\documentclass[12pt,preprint]{aastex}

\usepackage{epsfig}
\usepackage{amsmath}
\usepackage{amssymb}
\usepackage{amsbsy}
\usepackage{placeins}
\usepackage{color}

\newcommand{\D}{\mbox{d}}

\newcommand{\amr}{{\sc amr}}

\newcommand{\iles}{{\sc iles}}

\newcommand{\pdf}{{\sc pdf}}
\newcommand{\jpdf}{{\sc jpdf}}

\newcommand{\At}{\mathrm{At}}

\usepackage{ulem}

\newcommand{\gcc}{\mathrm{g~cm^{-3} }}
\newcommand{\cms}{\mathrm{cm~s^{-1}}}

\newcommand{\gtaprx}{\lower .1ex\hbox{\rlap{\raise .6ex\hbox{\hskip .3ex
        {\ifmmode{\scriptscriptstyle >}\else
                {$\scriptscriptstyle >$}\fi}}}
        \kern -.4ex{\ifmmode{\scriptscriptstyle \sim}\else
                {$\scriptscriptstyle\sim$}\fi}}}
\newcommand{\ltaprx}{\lower .1ex\hbox{\rlap{\raise .6ex\hbox{\hskip .3ex
        {\ifmmode{\scriptscriptstyle <}\else
                {$\scriptscriptstyle <$}\fi}}}
        \kern -.4ex{\ifmmode{\scriptscriptstyle \sim}\else
                {$\scriptscriptstyle\sim$}\fi}}}

\usepackage{epsf,color,dcolumn}

\newcolumntype{d}{D{.}{.}{-1}}

\epsfverbosetrue


\begin{document}

\title{Turbulence-Flame Interactions in Type Ia Supernovae}

\author{A.~J.~Aspden\altaffilmark{1}, J.~B.~Bell\altaffilmark{1},
  M.~S.~Day\altaffilmark{1}, S.~E.~Woosley\altaffilmark{2}, and M.~Zingale\altaffilmark{3}}

\altaffiltext{1}{Lawrence Berkeley National Laboratory, 1 Cyclotron
Road, MS 50A-1148, Berkeley, CA 94720}
\altaffiltext{2}{Department of Astronomy and Astrophysics, University
of California at Santa Cruz, Santa Cruz, CA 95064}
\altaffiltext{3}{Department of Physics and Astronomy, Stony
Brook University, Stony Brook, NY 11794}

\begin{abstract}
The large range of time and length scales involved in type Ia
supernovae (SN Ia) requires the use of flame models.
As a prelude to exploring various options for flame models, we consider,
in this paper,
high-resolution three-dimensional simulations of the
small-scale dynamics of nuclear flames in the supernova
environment in which the details of the flame structure are fully resolved.
The range of densities examined, 1 to $8 \times 10^7$~g~cm$^{-3}$, spans the
transition from the laminar flamelet regime to 
the distributed burning regime where small scale turbulence disrupts
the flame. The use of a low Mach number algorithm facilitates the
accurate resolution of the thermal structure of the flame and the
inviscid turbulent kinetic energy cascade, while implicitly
incorporating kinetic energy dissipation at the grid-scale cutoff.
For an assumed background of isotropic Kolmogorov turbulence with an
energy characteristic of SN Ia, we find a transition density between 1
and $3 \times 10^7$~g~cm$^{-3}$ where the nature of the burning
changes qualitatively. By $1 \times 10^7$~g~cm$^{-3}$, energy
diffusion by conduction and radiation is exceeded, on the flame scale,
by turbulent advection. As a result, the effective Lewis Number
approaches unity.
That is, the flame resembles a laminar flame, but is
turbulently broadened with an effective diffusion coefficient,
$D_T \sim u' l$, where
$u'$ is the turbulent intensity and $l$ is the integral scale.
For
the larger integral scales characteristic of a real supernova, the
flame structure is predicted to become complex and unsteady.
Implications for a possible transition to detonation are discussed. 
\end{abstract}

\section{INTRODUCTION}
\label{sec:intro}

The complex small-scale dynamics of turbulent thermonuclear flames are
essential to understanding SN Ia.  Since the range of length scales is
so large (star size $\sim 10^8$ cm; Kolmogorov scale and flame
thickness $< 1$ mm), full star calculations must use a subgrid
model to describe the burning on the unresolved scales. Turbulent
flame models \citep{niemeyerhillebrandt1995b} that move the flame at
the turbulent velocity have been successful at producing explosions
\citep{roepke2005} with pure deflagrations.  Another popular flame
model \citep{khokhlov1991} moves the flame at the speed dictated by
the Rayleigh-Taylor instability, and has also produced successful
explosions \citep{gamezo2003}, although the authors argue that a more
realistic supernova is produced when a detonation ensues late in the
explosion \citep{gamezo2005}.  These two flame models can produce
large differences in the flame speeds at small scales,
which will have a large
effect on the outcome of full-star simulations.  Validation of these models against resolved
flame calculations is needed to resolve this discrepancy.

As the flame propagates outward from the center of the star, it
encounters lower densities, the speed decreases, and the flame
thickens. A critical length scale in turbulent combustion is the
Gibson scale -- the length scale at
which the laminar flame speed equals the turbulent velocity (see for example \citealt{Peters00}),
\begin{equation}
l_G = l \left ( \frac{s_L}{u^\prime} \right )^3 \enskip .
\label{Eq:Gibson}
\end{equation}
Here $s_L$ is the laminar flame speed, and $u^\prime$ and $l$ are velocity and
integral
length scales characterizing the turbulence, and we have assumed Kolmogorov
scaling.  At a density of around $10^7~\gcc$, the flame becomes 
thick enough that turbulent eddies can disrupt its structure before
they burn away \citep{niemeyerwoosley1997}, that is, the flame
thickness is larger than the Gibson scale.  At this point, the burning
fundamentally changes character and the flame is said to be in the
distributed burning regime (see \citealt{Peters00} for some
discussion).

Here we look at the interaction of the flame and turbulence on the
scale of the flame, with the aim of validating turbulent flame models
and better understanding the transition to distributed burning.
Previous studies of flames interacting with turbulence have focused on
flames interacting with vortical flow (two-dimensional simulations
presented in \citealt{roepke-vort}) or statistical methods using
one-dimensional turbulence \citep{lisewski-ddt, lisewski-distributed}.
An open question is whether in the distributed burning
regime, a mixed region of partially burned fuel and ash can grow large
enough such that it can ignite a detonation
\citep{niemeyerwoosley1997,khokhlov1997,niemeyer1999}.
Detonations have gained
renewed interest lately as a means to burn the carbon/oxygen fuel left
behind near the center of the white dwarf in pure deflagration models
\citep{gamezo2005}.  Although, it is unclear whether detonations can
traverse the pockets of partially burned fuel \citep{golombek2005}.
Recent results also suggest that the size of the region of partially 
burned fuel needed to initiate a detonation is larger than previously
believed \citep{dursitimmes}. 
\citet{2008arXiv0803.1689P} 
discusses the role of turbulent intermittency on 
the conditions needed for transition to detonation. 

Previously, we showed the transition to the distributed burning regime
in two dimensions \citep{SNrt}, where the Rayleigh-Taylor instability
(RT), growing on scales smaller than the flame thickness itself, is
responsible for the transition.  
It has been suggested \citep{niemeyerkerstein1997} that turbulence generated
by RT in type Ia supernovae should obey Bolgiano-Obukhov (BO)
statistics -- formulated from considering a potential energy cascade.
However, recently is was shown that BO scaling only applies in
two-dimensions \citep{chertkov2003,scidac}.  
Three-dimensional RT-unstable flame calculations
have shown that turbulence
is indeed Kolmogorov in nature and becomes isotropic on the small scales
\citep{SNrt3d,cabotcook2006}.  As first discussed
in \citet{niemeyerkerstein1997}, BO scaling leads to a lower
transition to distributed burning, and therefore, makes a detonation
transition more difficult.  Therefore, although our two-dimensional
results led us to conclude that a DDT was unlikely \citep{SNrt}, a
three-dimensional study is warranted.

Figure~\ref{fig:regimes} shows how the relevant length scales vary with fuel
density.  In particular, the two red curves compare the laminar flame width
and the Gibson scale.  The laminar flame widths have been calculated from 
simulated one-dimensional profiles.  The Gibson scale is evaluated assuming a
turbulent intensity, $u^*$, of $10^7~\cms$ on a length scale, $L^*$, of
$10^6$~cm, as in \citet{niemeyerwoosley1997}.
We see that the transition to
distributed burning, assuming Kolmogorov turbulence, should occur around 
$3\times 10^7~\gcc$.   

In this paper, we present a three-dimensional study of turbulent thermonuclear
flames that explores a range of conditions from the flamelet regime to the
distributed burning regime.  Specifically, we use a flame sheet embedded in a
maintained turbulent velocity field.  Of particular interest is the response
of the flames to wrinkling by the turbulence, and the scaling of  the
turbulent flame speed with the turbulent intensity.  Thermal diffusion is 
orders of magnitude greater than species diffusion (the Lewis number is large)
\citep{timmeswoosley1992}
and has a significant effect on the turbulence-flame interactions.  Attention
is then focused on the distributed burning regime and potential implications
for a transition to detonation are discussed.

\section{SIMULATION DESCRIPTION}
\label{Sec:Numerics}

We use a low Mach number hydrodynamics code, adapted to the study of
thermonuclear flames, as described in \citet{SNeCodePaper}.  The
advantage of this method is that sound waves are filtered out
analytically, so the time step is set by the the bulk fluid velocity
and not the sound speed.  This is an enormous efficiency gain for low
speed flames.  The input physics used in the present simulations is
largely unchanged, with the exception of the addition of Coulomb
screening, taken from the Kepler code \citep{weaver:1978}, to the
$^{12}$C($^{12}$C,$\gamma$)$^{24}$Mg reaction rate.  This yields a
small enhancement to the flame speed, and is included for
completeness.  The conductivities are those reported in
\citet{timmes_he_flames:2000}, and the equation of state is the
Helmholtz free-energy based general stellar EOS described in
\citet{timmes_swesty:2000}.  We note that we do not utilize the
Coulomb corrections to the electron gas in the general EOS, as these
are expected to be minor in the conditions considered. 

The non-oscillatory finite-volume scheme employed here permits the use of
implicit large eddy simulation (\iles).  This technique captures the inviscid cascade
of kinetic energy through the inertial range, while the numerical error acts
in a way that emulates the physical effects of the dynamics at the grid
scale, without the expense of resolving the entire dissipation subrange.  The
approach was introduced by \cite{Boris92}, and has 
since been used by many authors (e.g.~\cite{Youngs91}, \cite{Porter92},
\cite{Grinstein04}, and \cite{Margolin06}).  An overview of the technique can be
found in \cite{GrinsteinBook07}.  \cite{Aspden08}
presented a detailed study of the technique using the present numerical
scheme, including a characterization that allowed for an effective viscosity
to be derived.  Thermal diffusion plays a significant
role in the flame dynamics, and so is explicitly included in the model, whereas species
diffusion is significantly smaller, and so is not explicitly included.

\begin{figure}
\centering
\epsscale{0.6}
\plotone{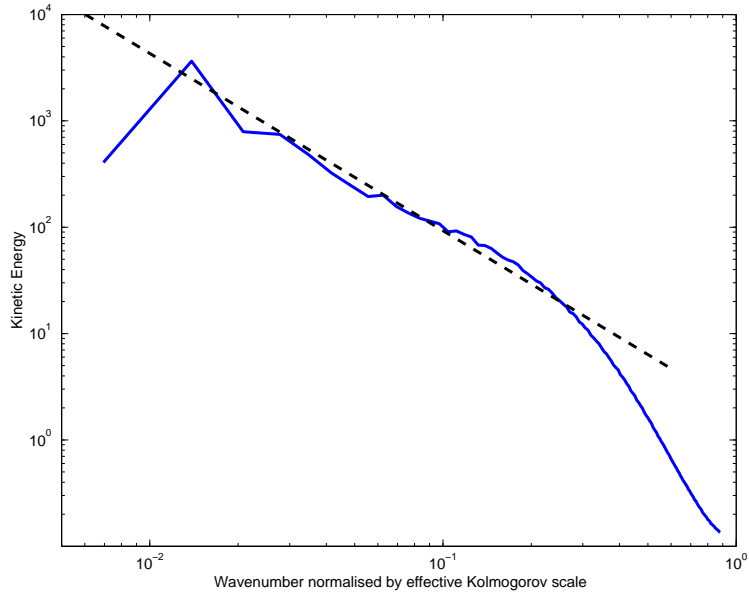}
\epsscale{1.0}
\caption{Kinetic energy wavenumber spectrum from a homogeneous isotropic
  turbulence simulation at $256^3$, taken from \cite{Aspden08}.  The same
  forcing term used to maintain the turbulence is used in the
  present flame calculations.  This figure demonstrates the range of scales
  captured by the \iles\ method for turbulent flow simulations at the
  resolution used in the present study.  The dashed black line denotes the
  minus five-thirds decay expected in an inertial range.} 
\label{Fig:Spectrum}
\end{figure}

The turbulent velocity field was maintained using a forcing term similar to that
used in the study of \cite{Aspden08}.  Specifically, a forcing 
term was included in the momentum equations consisting of a superposition of
long wavelength Fourier modes with random amplitudes and phases. The forcing term
is scaled by density so that the forcing is somewhat reduced in the ash. This
approach provides a way to embed the flame in a zero-mean turbulent background,
mimicking the much larger inertial range that these flames would experience in
a type Ia supernova, without the need to resolve the large-scale convective
motions that drive the turbulent energy cascade.  Figure~\ref{Fig:Spectrum}
shows an example kinetic energy wavenumber spectrum taken from \cite{Aspden08} at
a resolution comparable to that used here; the dashed line denotes a minus
five-thirds decay, and is illustrative of the turbulence found in the present
study.  \cite{Aspden08} demonstrated that the effective
Kolmogorov length scale is approximately $0.28\Delta x$, and the integral
length scale is approximately a tenth of the domain width.  

\begin{figure}
\centering
\epsscale{0.8}
\plotone{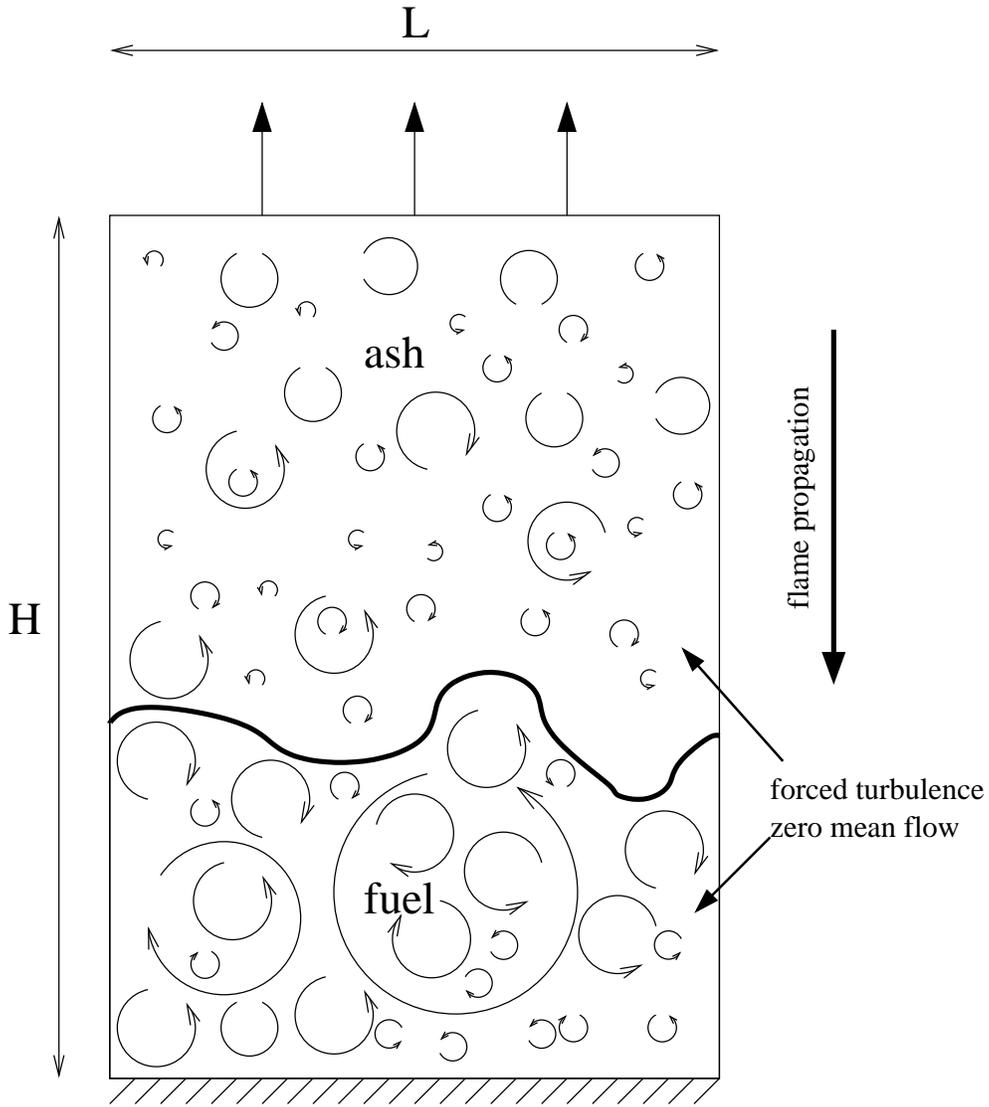}
\epsscale{1.0}
\caption{Diagram of the simulation setup (shown in
two-dimensions for clarity).
The domain is initialized with a turbulent flow and a flame is introduced into the
domain, oriented to that the flame propagates toward the lower boundary.
The turbulence is maintained by adding a forcing term to the momentum equations.
The top and bottom boundaries are outflow and solid wall boundaries, respectively.
The side boundaries are
periodic.}
\label{Fig:Setup}
\end{figure}

Figure~\ref{Fig:Setup} shows the simulation setup.  The simulations were
initialized with carbon fuel in the lower part of the domain and magnesium ash
in the upper, resulting in a downward propagating flame.  A high-aspect ratio
domain was used to allow the flame sufficient space to propagate.  Periodic
boundary conditions were prescribed laterally, along with a free-slip base,
and outflow at the upper boundary.

Five simulations were run to investigate the turbulence-flame interactions in
different burning regimes.  The five cases will be referred to as cases (a)
through (e).  The study was designed so that $l_G/l_L$ varied across a number
of orders of magnitude, from approximately $4\times10^3$ in 
case (a) corresponding to the flamelet regime, to $2\times10^{-5}$ in case (e)
corresponding to the distributed burning regime.  The different burning
regimes were achieved by varying the density of the carbon fuel from
$8\times10^7$ for case (a) down to $1\times10^7$ in case (e).  Case (a)
corresponds to near the center of a supernova, has a thin laminar flame and
relatively-low turbulence levels.  Case (e) corresponds to the conditions near
the edge of a supernova, has a thicker laminar flame and relatively more
intense turbulence.  The derivation of the turbulence conditions will be
described below.

\begin{figure}
\begin{center}
\plotone{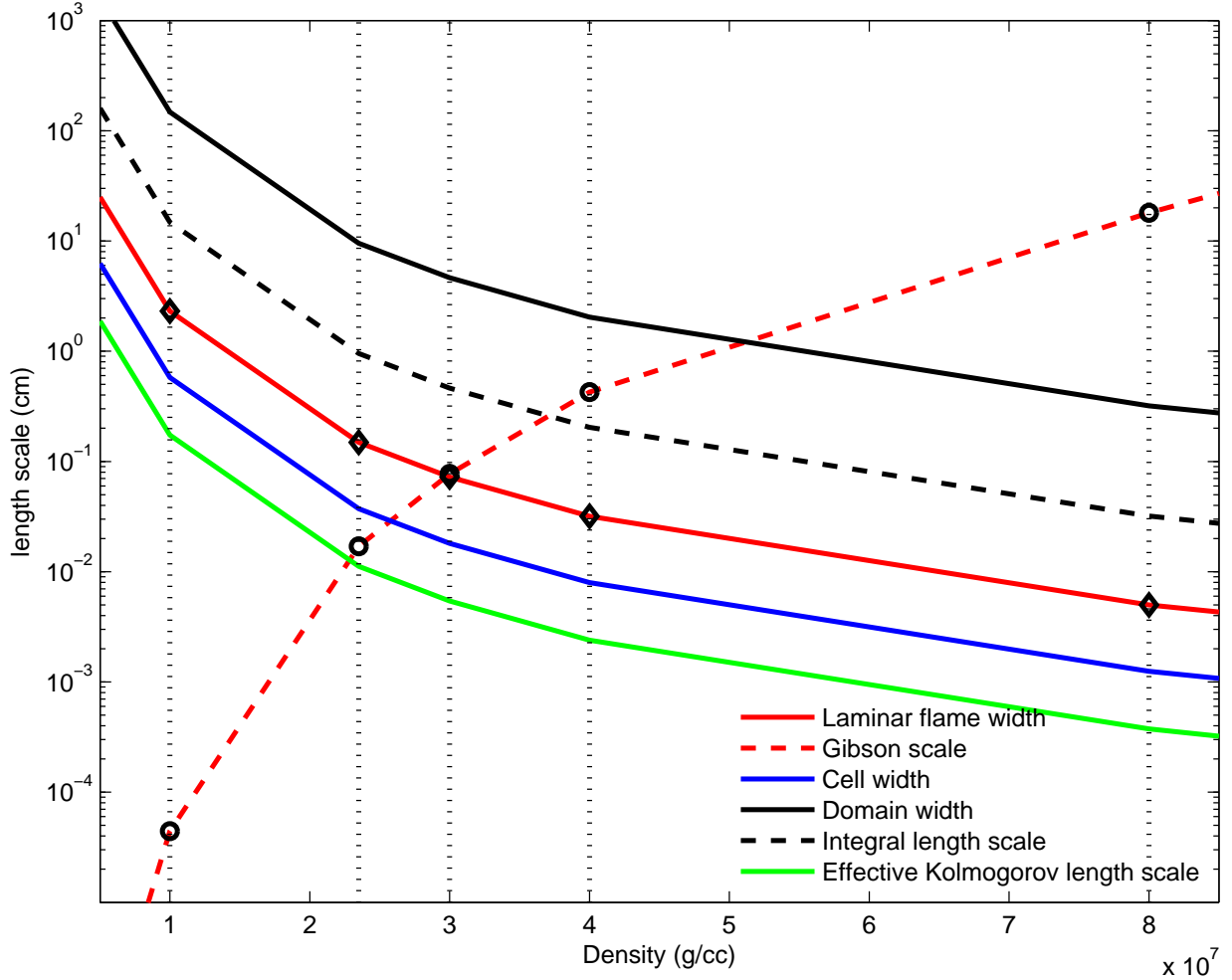}
\end{center}
\caption{Comparison of the variation of length scales at different
fuel densities.  The solid red line shows how the laminar flame width increases with
decreasing density.  The dashed red line shows how the Gibson scale ($l_G$)
decreases with density (Kolmogorov scaling appropriate for a star has been
assumed, see equation~\ref{Eq:Gibson}).  The flame thickness and Gibson scale are 
approximately equal at a density of $\rho_7\approx3$.  This marks the
transition between the flamelet and distributed burning regimes.  The cell width is shown
by the solid blue line; the resolution of the turbulent simulations was chosen
such that $l_L=4\Delta x$.  The available computational resources limit the
size of calculation and so the domain width in each case was $L=256\Delta x$
and is shown by the solid black line.  The forcing term used to maintain the
turbulence imposes an integral length scale that is approximately a tenth of
the domain width, and is shown by the dashed black line.  \cite{Aspden08}
demonstrated that the effective Kolmogorov length scale using this numerical
scheme is approximately $0.28\Delta x$, and is shown by the solid green line.
The vertical dashed lines denote the densities of the five turbulent simulations.}
\label{fig:regimes}
\end{figure}

\begin{figure}
\centering
\epsscale{0.6}
\plotone{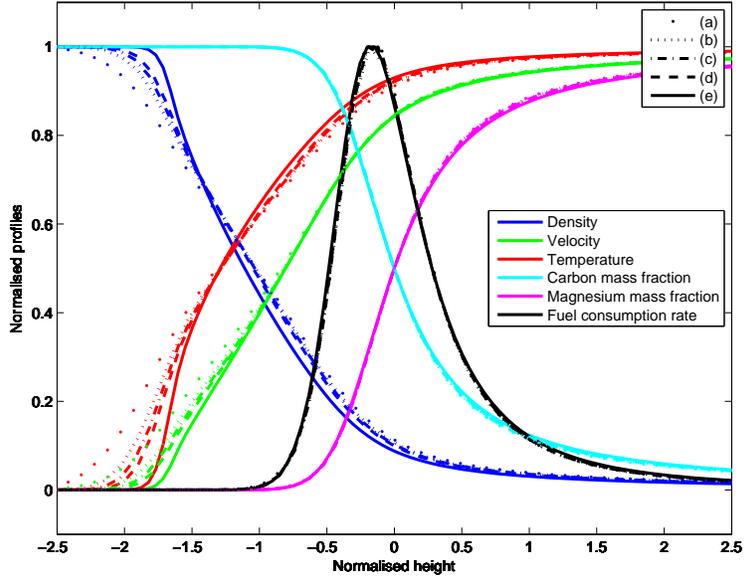}
\epsscale{1.0}
\caption{Normalized laminar flame profiles.  Each color denotes a different
  quantity, as indicated by the legend.  The line style corresponds to the
  five different cases; the dotted line is case (a) through to the solid line
  for case (e).  The length scales have been normalized by the laminar flame
  width, defined to be the carbon width $l_L=(\Delta X_C)/\max|\nabla
  X_C|$, where $X_C$ is the carbon mass fraction.  Each quantity has been
  normalized by the minimum and maximum values attained (see
  table~\ref{Tab:SimProperties}),
  $\bar{q}=(q(z)-q_{\min})/(q_{\max}-q_{\min})$.  Note the velocity is given
  in the frame of reference of the flame.} 
\label{Fig:LaminarProfiles}
\end{figure}

For each case, a flat laminar flame simulation was first run to steady-state
to establish a laminar flame width and speed.  The laminar flame width was
taken to be the carbon width, specifically
$l_L=(\Delta X_C)/\max|\nabla X_C|$, where $X_C$ is the carbon mass
fraction.  The variation of laminar flame thickness with fuel density is shown
by the solid red curve in figure~\ref{fig:regimes}.  The laminar flame
solution also provides initial flame conditions for the turbulent simulations.  
Figure~\ref{Fig:LaminarProfiles} shows normalized laminar profiles for the
five cases.  The length scale has been normalized by the carbon width, and
each quantity has been normalized by the minimum and maximum value across the
domain,  i.e.~$\bar{q}=(q-q_{\min})/(q_{\max}-q_{\min})$ for a generic
quantity $q$.
Properties of the laminar flames are summarized in Table \ref{tbl:1}.

Based on the resolution studies presented in \cite{SNeCodePaper, SNrt},
the resolution for the turbulent flame simulations was chosen to have four
cells across the laminar flame width.  Specifically, the carbon profile 
from each laminar flame simulation was measured, $l_L$, and $\Delta x$
set to $l_L/4$, which is shown by the solid blue line in
figure~\ref{fig:regimes}, and, following \cite{Aspden08}, the effective
Kolmogorov length scale is shown by the solid green line.  This choice of cell
width corresponds to many more cells across the
entire flame; at the resolution of the turbulent flame simulations to follow, the
spatial extent shown in figure~\ref{Fig:LaminarProfiles} would be spanned by
twenty computational cells.  Since it is expected that the turbulence will
thicken the flame, this choice of resolution is believed to be sufficient to
resolve the reaction zone of the turbulent flame.

Computational expense restricts the domain to a cross-section of $256$ cells,
and so, since $\Delta x$ is known, the domain width follows, and the forcing
term determines the integral length scale.  The domain width and integral
length scales are shown in figure~\ref{fig:regimes} by the solid and dashed
black lines, respectively.  It should be emphasized that the integral length
scale and effective Kolmogorov length scales are restricted by computational
expense and are not reflective of the values found in a supernova.

Turbulence characteristics can be estimated using a simple
representation shear generation by the
Rayleigh-Taylor instability.  On the large scale of the supernova, the
flame will be unstable with a typical feature size of $L^* \sim
10^6$~cm.  The velocity on this scale can be estimated using the
Sharp-Wheeler model \citep{sharp1984},
\begin{equation}
u^*(L^*) \sim \frac{1}{2} \sqrt{\At g L^*} \enskip .
\end{equation}
For an Atwood number, $\At \sim 0.5$ and gravitational acceleration,
$g \sim 10^9~\mathrm{cm}~\mathrm{s}^{-2}$, one has $u^*(L^*) \sim
10^7~\cms$, in agreement with the numbers used in
\citet{niemeyerwoosley1997}. More detailed simulations by
\citet{Roepke2007} in 3D find a distribution of turbulent intensities 
peaking at 10$^7$~cm~s$^{-1}$, but with a high-velocity tail extending
out to 10$^8$ cm s$^{-1}$.  For the current simulations, the turbulent
kinetic energy released by the large-scale RT motions is assumed to cascade
down to the flame scale following a Kolmogorov spectrum, as
\begin{equation}
u^\prime(l) = u^*(L^*) \left (\frac{l}{L^*}\right)^{1/3} \enskip ,
\end{equation}
therefore, given the integral length scale, $l$, the turbulent
velocity, $u^\prime(l)$, can be computed. Throughout this paper, $u^*$
= 10$^7$ cm s$^{-1}$ and $L^*$ = 10$^6$ cm have been assumed.

In cases (a)-(d),  the aspect ratio of the domain was 4:1 and the flame placed
one domain width from the upper boundary.  In case (e) an aspect ratio of 8:1
was required to accommodate the flame in the distributed burning regime, and
the flame was initially placed at the midpoint of the domain.  In each case,
adaptive mesh refinement was used to reduce the computational expense.  A
base grid with 128 cells across was used, with one level of \amr\ with 
refinement factor 2 concentrated around the flame sheet.  This gave an
effective resolution of $256\times256\times1024$ for cases (a)-(d) and
$256\times256\times2048$ for case (e).
Experiments with additional resolution confirmed the adequacy of this
choice.  It was also ensured that the use of \amr\ had no detrimental
effect on the turbulence.  Specifically, the energy containing scales
were well captured by the base grid (see \cite{Aspden08} for a study
of maintained homogeneous isotropic turbulence using the same
approach) and there was adequate refinement ahead of the flame to
ensure that the cascade of energy to smaller scales in the refined
region (which have a short turn over time) could occur before these
scales could interact with the flame.
Table~\ref{Tab:SimProperties} summarizes the properties of the turbulence simulations.

\section{RESULTS}
\label{Sec:Results}

\begin{figure}
\centering
\plotone{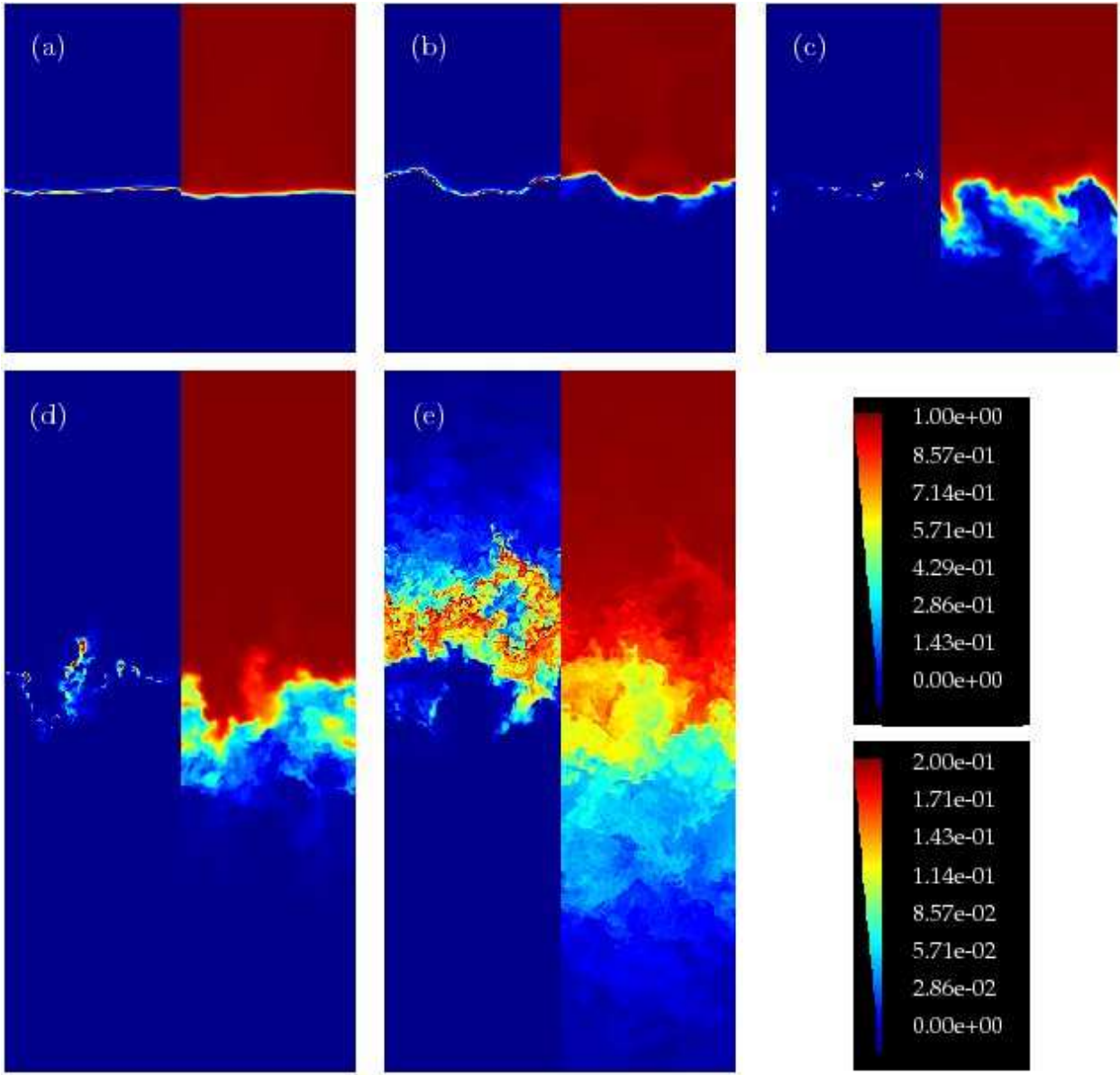}
\caption{Instantaneous vertical slices through each flame.  In each case, the
  left-hand panel is fuel consumption rate, normalized by the corresponding
  laminar value (except for case (e), which was
  normalized by one fifth of the laminar value because it burns much less
  intensely than the laminar flame), and the right-hand panel is the temperature field, again
  normalized by the laminar value, (refer to table~\ref{Tab:SimProperties} for
  the values).  The top legend shows the range for each normalized value
  except fuel consumption rate in case (e), which is shown by the lower
  legend.}
\label{Fig:Slices}
\end{figure}

In each case, the flow undergoes an initial transition, as the turbulence
wrinkles and thickens the flame, until a quasi-steady state is established.
Figure~\ref{Fig:Slices} shows instantaneous vertical slices of fuel
consumption rate (left panel) and temperature (right panel) for each of the
five cases once the quasi-steady state has been reached.  In each case, the
values have been normalized by the corresponding laminar values, with the
exception of case (e) where the local fuel consumption rate is much lower and
so has been normalized by a fifth of the laminar value for contrast.  By
construction, the length scales are normalized by the laminar flame width.
Half of the domain is shown for cases (a)-(c) and case (e), and the entire
domain is shown for case (d).

Case (a) presents smooth and even burning, and is perturbed very little by the
background turbulence.  The temperature profile remains sharp.  In case (b), the flame
surface has been deformed by the turbulence.  Both regions of enhanced burning
and regions of decreased burning are observed, and appear to be correlated
with the curvature of the flame sheet.  Specifically, enhanced burning appears
to occur where the center of curvature is within the fuel, and decreased when
the center of curvature is in the ash.
The temperature field presents regions that are sharp, and regions that appear
to be more diffuse.  Again, this appears to be correlated with curvature,
with the more diffusive regions occurring where the center of curvature is in
the products.

In cases (c) and (d), as $l_G/l_L$ decreases further, the background turbulence
becomes increasingly influential; the temperature field becomes more
mixed, and the deformation of the flame surface increases.  The burning
appears to occur in small high-intensity pockets, punctuated by regions of
local extinction.

In case (e), a dramatically different burning mode is observed.  The
temperature mixed region and the burning region are much broader.  The
burning appears to be much less intense (recall the image has been normalized
by a fifth of the laminar value for contrast) and is restricted to the high
temperature end of the mixing zone.  There is no well-defined flame
surface, but a broad flame brush.  Interestingly, there appears to be some
residual low-level burning well above the main burning region, suggestive of
incomplete burning in the main flame zone.

\begin{figure}
\centering
\plotone{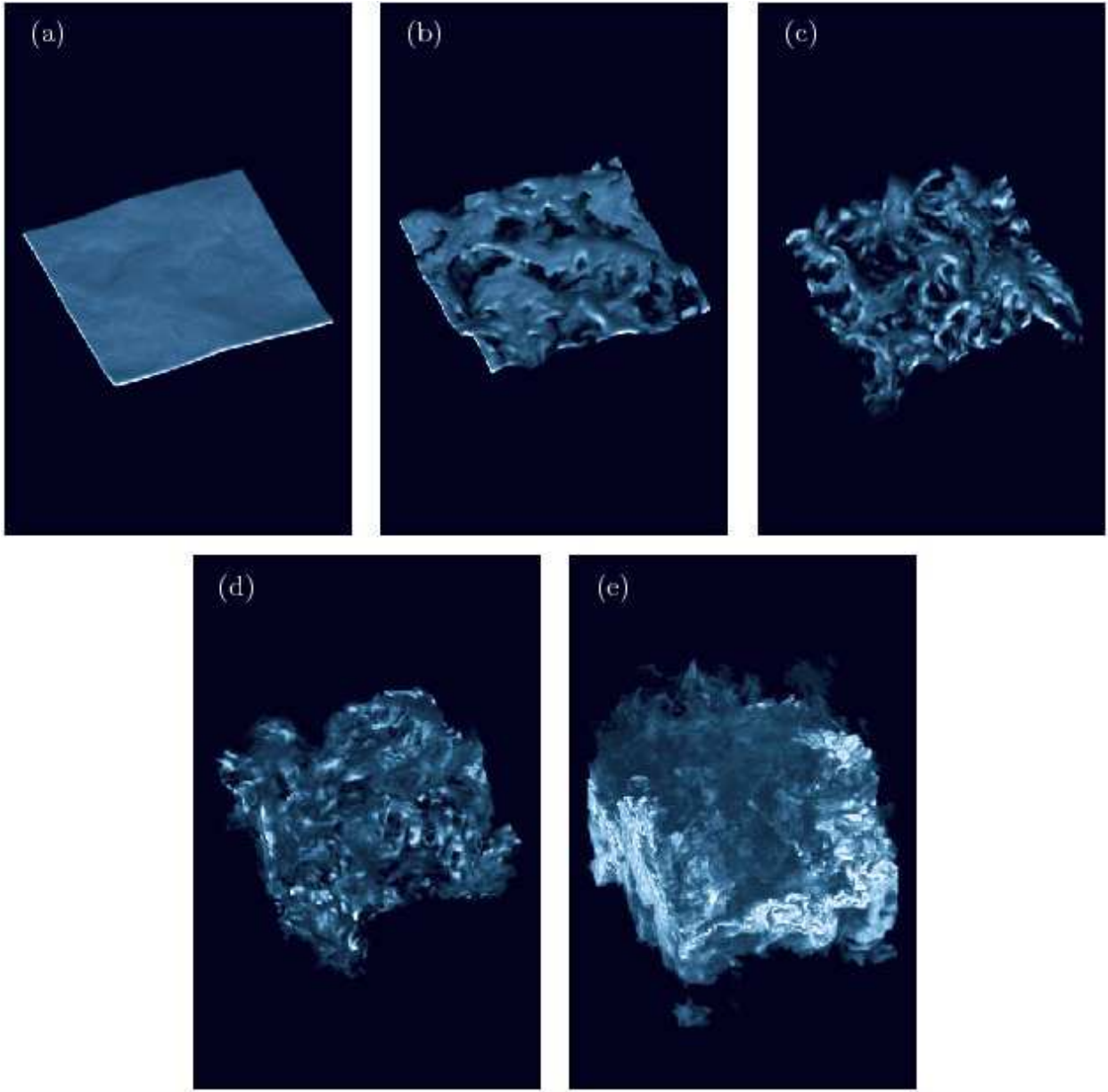}
\caption{Three dimensional instantaneous fuel consumption rate in each
  of the five cases, rendered such that the burning intensity is
  linearly related to opacity of the image.
  In the flamelet regime, the flame is coherent and
  flat.  As the relative turbulence increases, the flame becomes increasingly
  disrupted, which leads to regions of enhanced burning and regions of local
  extinction.  Finally, in the distributed burning regime in case (e), the
  character of the burning changes dramatically; the flame is greatly
  broadened, burns much less intensely (recall that the normalization
  is by a fifth of the laminar value), but the overall fuel consumption rate
  is enhanced.}
\label{Fig:3dRenders}
\end{figure}

These observations are further reinforced by three-dimensional renderings of
the fuel consumption rate, shown in figure \ref{Fig:3dRenders}, which
elucidates the three-dimensional structure of the flames.  The images
have been scaled in the same way as the corresponding slices.  Case (a) burns
as a coherent connected flame sheet, but as the relative turbulence level
increases in cases (b) to (d), the flame sheet becomes increasingly disrupted
and presents regions of local extinction.  Finally in case (e), in the
distributed burning regime, the behavior is completely different with a flame
brush that is much broader but burns less intensely than the laminar flame.

The curvature effects observed and the resulting burning rates are a
consequence of the thermo-diffusively-stable nature of the flames; the Lewis
number is high -- thermal diffusion is much greater than species
diffusion.  Where the background turbulence elevates part of the flame
surface, resulting in negative curvature, there is a focusing of heat by
diffusion.  Consequently, as the reaction rate is extremely sensitive to
temperature, the fuel burns quickly, and the flame is flattened.  Conversely,
where the turbulence pushes the flame downwards, creating positive curvature,
temperature diffusion leads to a defocusing of heat, and the burning rate
decreases.  Again, this tends to flatten the flame.  Curvature effects will be
explored further below.

The instantaneous global turbulent flame speed normalized by the laminar flame
speeds are shown in figure~\ref{Fig:AllSpeeds}; the time scale is normalized
by the laminar burning time -- the time it takes the laminar flame to burn one flame width.  Note
the different time axis for case (e).  Case (a)
burns almost exactly at the the laminar flame speed.  Curiously,
cases (b) and (c) actually burn more slowly overall than their laminar
counterparts, and case (d) is only around 30\% faster.  However, case (e)
burns between approximately five and six times the laminar flame speed.  

To investigate the effects of curvature, it is useful to be able to define a
flame surface.  There are a number of ways to do this, but it has been found
that using isosurfaces of temperature or fuel mass fraction is a practical way
that avoids the difficulties associated with the local extinction in these particular
flames.  We choose values for the isosurfaces based on the temperature or fuel mass
fraction values at the peak local fuel consumption rate in
the laminar flame.
The isosurface
was then located by using a standard ``marching cubes'' algorithm, and
optimized using the ``QSlim'' algorithm, see~\cite{Garland99}.  Since the temperature field is more
diffusive than the fuel, the resulting isosurface was found to be much
smoother for temperature than the carbon, particularly where the isosurface
was within the fuel itself.  Furthermore, because of the higher levels of
turbulence in cases (d) and (e), the isosurfaces were found to be too
contorted to be useful.

Since the fuel consumption occurs over a finite width and is not localized to the flame
surface, a local consumption-based flame speed was evaluated by integrating
the fuel consumption rate through the surface.
This involved constructing a
set of integral curves along the gradient vector of the progress variable
through each point on the isosurface, ensuring that the curves extended well
beyond the region where the fuel consumption rate decreases to zero.
(We note that for this construction, we orient the normal to point into the ash.
With this orientation, the curvature, $\kappa$ is negative when the center
of curvature is in the fuel.)
These
integral curves provided bounding edges of prism-shaped subvolumes that
effectively cover the reaction zone.  The consumption-based flame
propagation speed was then computed over each subvolume of the reaction zone
by integrating the computed fuel consumption rate over the subvolume.  
The local consumption-based flame speed was defined as
\begin{equation}
s^{l}_T = \frac{1}{\rho_0 X_{F,0} A}
\int_{\Omega} \rho \dot{\omega}_F \,\D{\Omega}
\label{Eq:LocalIntFCR}
\end{equation}
where $\rho_0 X_{F,0}$ is the initial fuel
density, $\Omega$ is the subvolume, $\rho \dot{\omega}_F$ is
the local fuel consumption rate, and $A$ is the area of intersection of the
flame with $\Omega$.  Defining the local speed in this way has the property
that the global burning speed is the integral of $s^{l}_T$ over the
isosurface.  For  additional detail about construction of the elements,
$\Omega$, see~\cite{BellEtal2005b}.

\begin{figure}
\centering
\epsscale{0.65}
\plotone{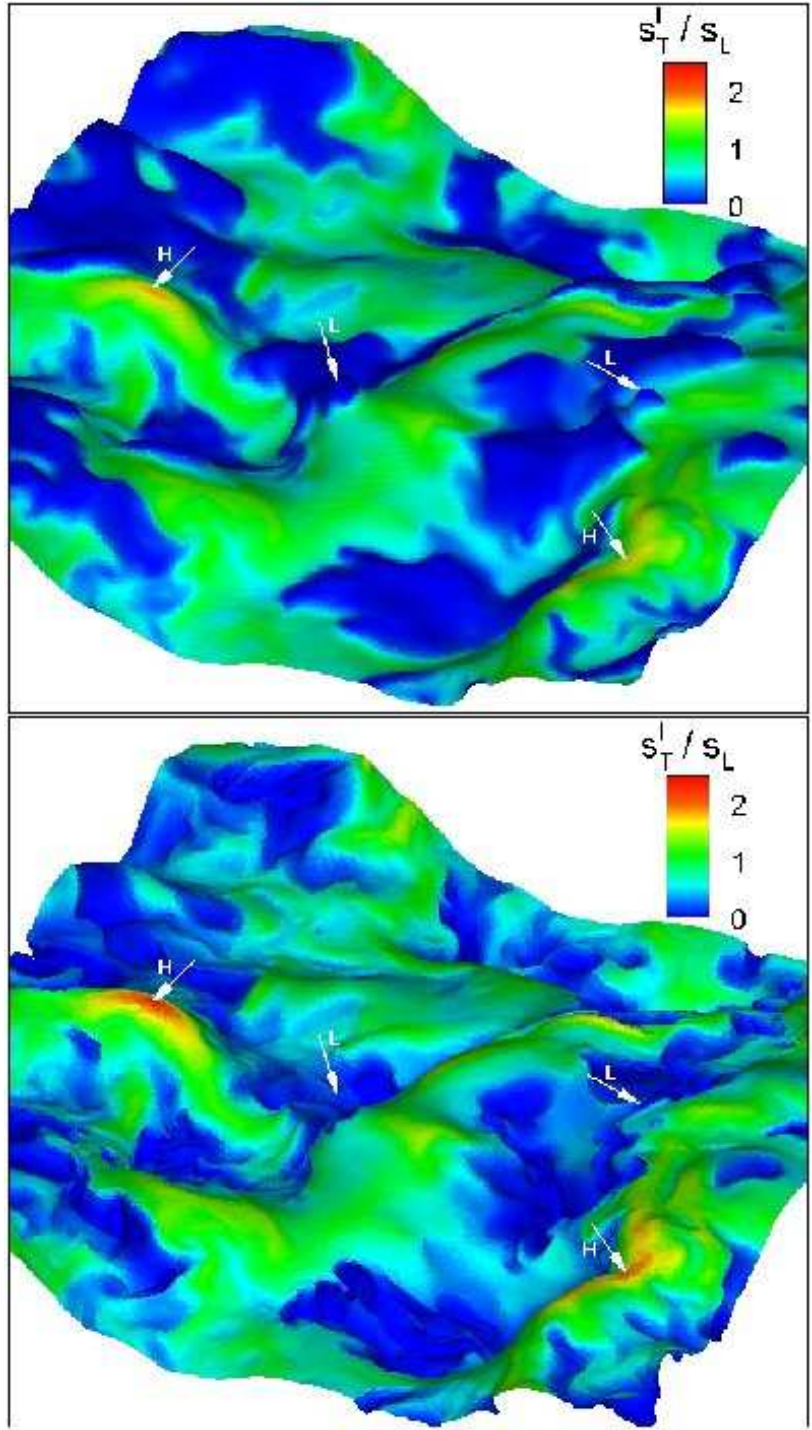}
\epsscale{1.0}
\caption{Comparison of isosurfaces based on temperature (top) and carbon mass
  fraction (bottom) for case (b).  The surfaces are colored by the locally
  integrated fuel consumption rate (see the text and
  equation~\ref{Eq:LocalIntFCR}).  The arrows labeled `H' and `L' highlight
  regions of high and low fuel consumption rate, respectively, both of which
  are correlated with negative curvature in the temperature isosurface.  However, in
  the carbon isosurface, the surface is much lower, indicating that although
  the fluid is at a sufficient temperature to burn, the fuel is
  absent.}
\label{Fig:SurfacesR4}
\end{figure}

Figure~\ref{Fig:SurfacesR4} compares instantaneous temperature and
carbon isosurfaces for case (b), and is colored by the integrated fuel
consumption rate, $s_T^l$.  The correlation between burning rate and curvature observed
in figure~\ref{Fig:Slices} is more apparent here.  Specifically, the regions
of high fuel consumption rate are strongly correlated with regions of negative
curvature, i.e.~where the centers of curvature are within the fuel.  Two
examples are highlighted by arrows labeled `H'.  This correlation appears to
differ slightly between the two isosurfaces.  On the temperature surface,
there are regions of negative curvature where the burning is very low; two
examples are shown by arrows labeled `L'.  However, the carbon surface does
not appear to pass through these regions; the labels are in the same place,
but the carbon surface is much lower.  The temperature and carbon mass
fraction have become decorrelated.  Regions exists around the peak
burning temperature where there is no fuel, and so burning cannot occur.
This is a direct result of the competition between mixing by turbulence and
thermal diffusion.

\begin{figure}
\centering
\plotone{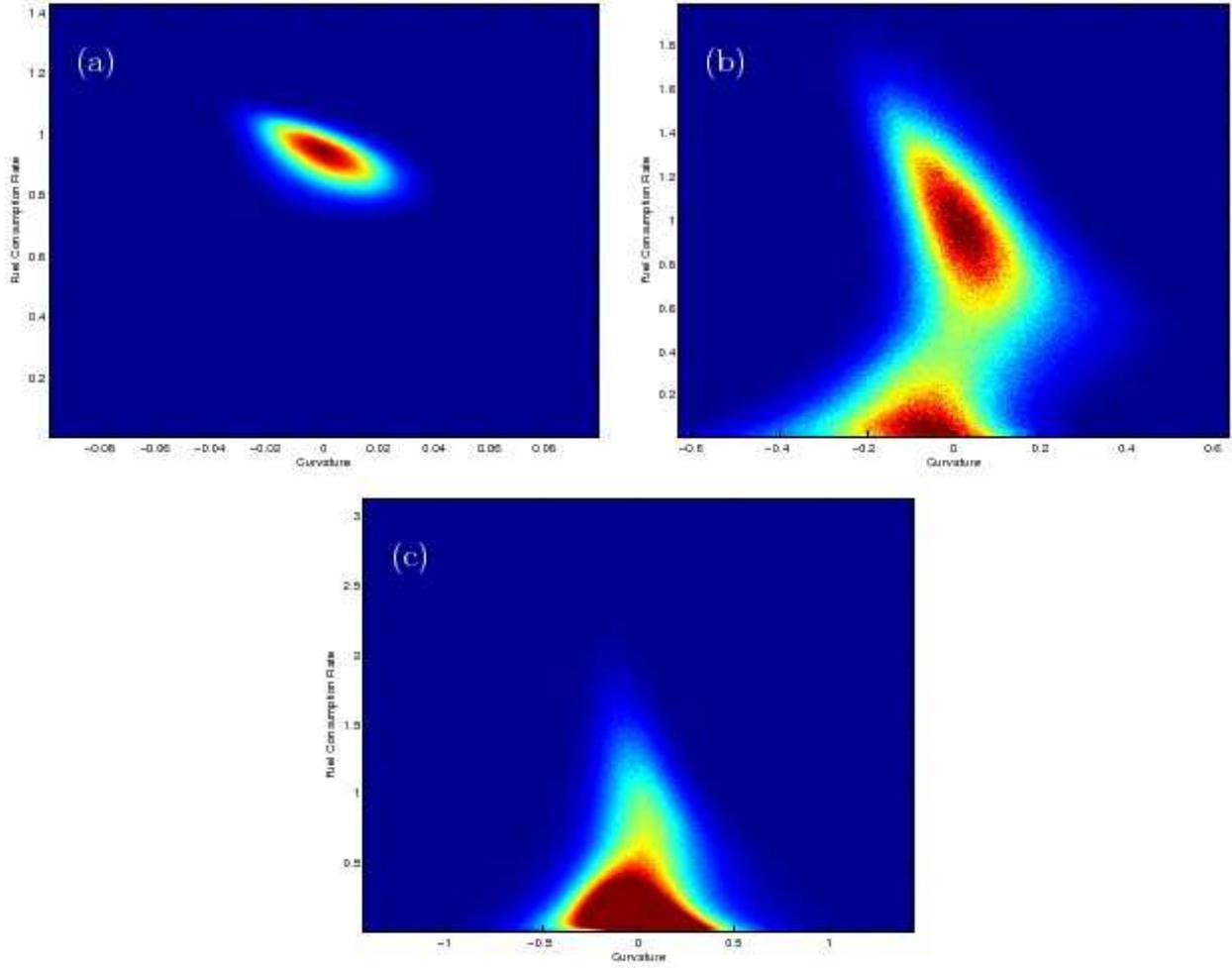}
\caption{Joint probability density functions of the locally integrated fuel
  consumption rate ($s^l_T$) normalised by the laminar flame speed ($s_L$)
  against curvature of the temperature isosurface ($\kappa_T$) normalized by
  the laminar flame width ($l_L$); (a) $\rho_7=8$, (b) $\rho_7=4$, (c)
  $\rho_7=3$.  The (negative) correlation is clear, specifically that the fuel
  consumption rate increases with negative curvature.  For case (b), the
  regions of negative curvature labeled `L' in figure~\ref{Fig:SurfacesR4},
  with low fuel consumption rate, explain the bimodal \pdf\ here.}
\label{Fig:PdfCurvFcr}
\end{figure}

Figure~\ref{Fig:PdfCurvFcr} shows joint probability density functions (\jpdf) of
curvature (normalized by the laminar flame width) and local integrated
fuel consumption rate (normalized by the laminar value), based on the
temperature isosurface, ensemble-averaged over a number of time-points after
the flames have reached a quasi-steady state.  The (negative) correlation is
clear and quantifies the between curvature and local flame speed observed above.
It is also evident that there is a significant change in burning
across the three cases. Case (a) shows that the majority of the fuel
consumption occurs around the laminar rate.  Case (b) demonstrates that the
flame is bimodal, in the sense that there are regions that are burning, which
are negatively correlated with curvature, but there are also regions of low
burning or local extinction, which occur with curvature of both signs.
This suggests that where the flame
is burning, it is burning close to the laminar value (with low probability
variability), but because there are also regions of local extinction, the
overall burning rate per unit area is lower than the laminar value.  Finally,
in case (c), there is no preferred rate of burning, a slight negative
correlation with curvature, and significant regions of local extinction.

\begin{figure}
\centering
\epsscale{0.6}
\plotone{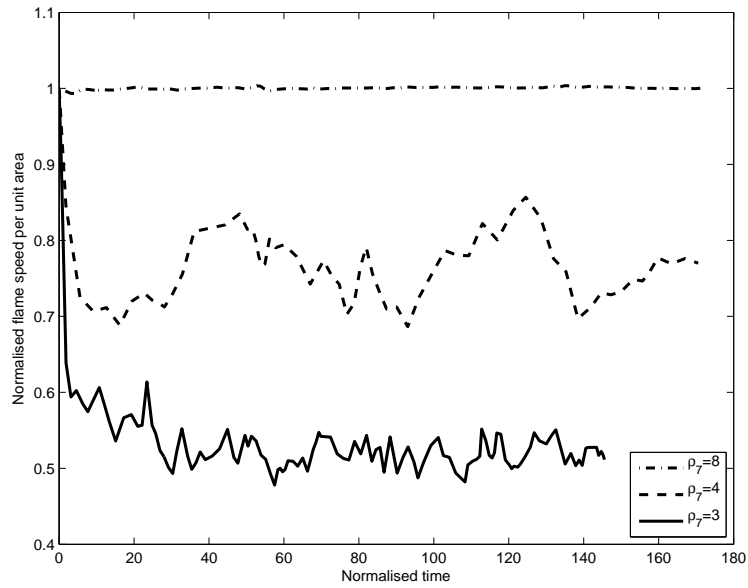}
\epsscale{1.0}
\caption{Normalized turbulent flame speed per unit area.  The turbulent flame
  speed has been normalized by the laminar flame speed, and the area has been
  taken to be the area of the temperature isosurface.  The time scale has
  been normalized by the laminar burning time ($l_L/s_L$).  This demonstrates
  that a Markstein correction is required to construct a flame
  model that captures the curvature dependence of the flame speed.}
\label{Fig:AreaSpeeds}
\end{figure}

A common approach used for determining a model flame speed involves assuming
that the flame is burning locally at the laminar flame speed, and that
global burning rate is equal to the laminar flame speed times the area of the
flame. Figure~\ref{Fig:AreaSpeeds} shows the normalized burning rate per unit
area for cases (a-c).  If the flame-modeling approach described above was
appropriate for these flames, this measure would be close to unity.  This is
indeed seen to be the case for case (a), as the flame is burning in a very
similar way to the laminar flame.  However, it is too simplistic an
approximation for other two cases.  The strong response to curvature modulates
the local burning speed.  The values of approximately $0.8$ and $0.5$ suggest
that using the laminar flame speed times the area of the flame would lead to
an overestimation of the global burning by factors of approximately $1.25$ and
$2$ respectively.
The analysis indicates that even in the flamelet regime, a Markstein correction
is needed to predict the local flame speed accurately.

\begin{figure}
\centering
\epsscale{0.9}
\plotone{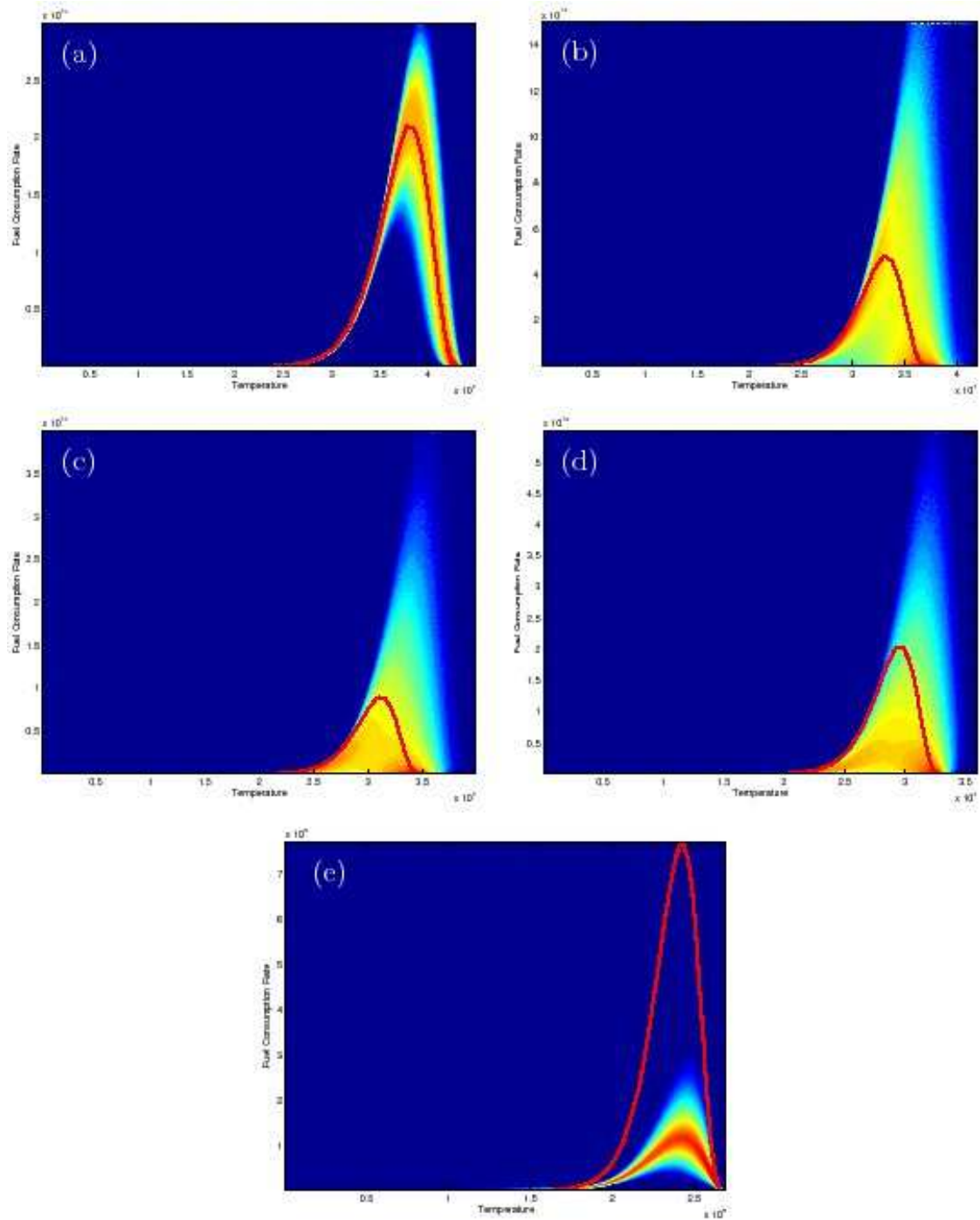}
\epsscale{1.0}
\caption{First moment with respect to fuel consumption rate of the joint
  probability density function of fuel consumption rate against temperature.
  Again, the red line denotes the laminar burning path, with which case (a) is
  in close agreement.  Case (e) has collapsed to a curve that is different
  from the laminar curve, and reiterates the lower local fuel consumption rate
  observed in this case.  The intermediate cases (b) through (d) show
  variation around the mean path and move between the laminar and turbulent
  cases, and it is particularly clear that greatly increased local fuel
  consumption rates are observed albeit with low probability.}
\label{Fig:PdfTempFcrM1}
\end{figure}

The way in which the burning occurs can be analyzed further by considering
other joint probability density functions.  Figure~\ref{Fig:PdfRhoXcTemp}
shows the \jpdf s for all five cases of temperature and carbon density,
and figure~\ref{Fig:PdfTempFcrM1} shows fuel consumption rate against
temperature.  Here the first moment has been taken with respect to fuel
consumption rate, specifically, for \jpdf\ $P(Q,T)$ for fuel consumption rate
$Q$ and temperature $T$, the first moment is defined as $QP(Q,T)$. This
highlights the burning regions and prevent the large 
proportion of the domain that is not burning from dominating the \jpdf.
Again, the \pdf s have been ensemble-averaged.  The
solid red line in each case denotes the laminar flame correlation.  In case
(a), it is clear that the burning follows an almost identical path to the
laminar flame.  Case (b) present a similar correlation, but there is greater
variability around the main burning path.  In cases (c) and (d), the main
burning path appears to have shifted away from the laminar flame, and there is
a large amount of variability from that main path.  In case (e), a dramatic
change is observed, the burning path has collapsed to a single path that is
significantly removed from the laminar flame.  In this case, the turbulent
mixing dominates the thermal diffusion and so the temperature and fuel cannot
become decorrelated.  Therefore, a single burning path is observed, which is
different than the laminar flame, with very little variability.

Figure~\ref{Fig:PdfXcFcrM1} shows the joint probability density function of
fuel consumption rate and carbon mass fraction for case (e), again the
first moment with respect to fuel consumption rate has been taken to highlight
the burning region.  Similar to the corresponding temperature plot, the
burning path has collapsed to a curve that is significantly different than the
laminar flame and there is little variability around it.  Importantly, this
figure demonstrates that the burning occurs at much lower mass fractions than
in the laminar flame.

A normalized instantaneous flame structure in case (e) is compared with the
laminar flame structure in figure~\ref{Fig:FlameStructure}.  The turbulent
flame structure has been obtained by taking planar averages.  Here, the length
scale has been normalized by the laminar flame carbon width, and each of the
other quantities has been normalized by the corresponding laminar values,
expect for the velocity, which has been normalized by the turbulent burning
speed due to its significantly enhanced value.  The turbulent structure is
dramatically different.  Note how the shape of the temperature
and density profiles have changed, becoming closer to hyperbolic
tangents.  In particular, the width of the temperature profile has
decreased relative to the carbon profile because the mixing is
dominated by the turbulence rather than thermal diffusion.  Therefore,
any comparison between the turbulent and laminar flame widths will
depend on the arbitrary choice of width definition.
In spite of this, a rough estimate of the
carbon width is approximately 60 times the laminar flame width (roughly 140 cm).  
Some effect of thermal diffusion is still evident as the
temperature profile and therefore density profile are slightly wider than the
species profile; the effective Lewis number is still greater than unity.  
One of the biggest differences between the two flame structures is the fuel
consumption rate.  In the turbulent case, the fuel consumption is very much
lower, the peak value is approximately an eighth of the laminar value.  The
physical location also appears to have shifted to a relatively higher $z$
location.  It should be noted that in the distributed burning regime,
the flame width is strongly affected by the integral length scale of
the turbulence (and hence the domain size of the present
calculations).  The effect of the integral length scale will be the
subject of future work, but can be predicted by Damk\"ohler scaling
\citep{Damkohler40}, which will be discussed in
section~\ref{sec:conclusions}.  Some preliminary calculations (see
\cite{WoosleySciDAC08} and \cite{Woo08}) have produced encouraging
results.

\section{CONCLUSIONS}
\label{sec:conclusions}

We have explored, in three dimensions, the properties of flames
interacting with isotropic Kolmogorov turbulence for the conditions
appropriate to a Type Ia supernova. In particular we have examined the
consequences of a turbulent energy cascade with $\varepsilon = u^{*3}/L^*
= 10^{15}$ erg g$^{-1}$ s$^{-1}$. This might correspond, for example,
to a macroscopic integral scale of 10 km in the supernova and speed on
that scale of 100 km s$^{-1}$.  The range of densities explored is
characteristic of the transition from the flamelet regime
to distributed burning at this energy density. In particular, the
Karlovitz number ($(l_L/l_G)^{1/2}$) varies from 0.017 at the highest
density to 230 at the lowest. Two regimes of burning are clearly
discernible. At the highest density, turbulence merely wrinkles an
otherwise laminar flame. The requirement that the flame be resolved
meant that we were not able to examine a sufficiently large integral
scale ($l \gg l_G$) to see the multiply-folded flames that should be
present in the flamelet regime. The width and speed of these
individual flamelets were not greatly affected by the turbulence.

As the density is decreased, the flames initially remain in the flamelet regime;
however, the increase in turbulence intensity relative to the flame speed leads to
enhanced wrinkling of the flame. In this regime, we see significant Lewis number effects.
The combination of the sensitivity of the nuclear reaction rates to temperature and the
large Lewis number leads to significant variability in the local burning rate along the flame
surface, with regions of negative curvature burning much more intensely than regions
of positive curvature.

\begin{figure}
\centering
\epsscale{0.9}
\plotone{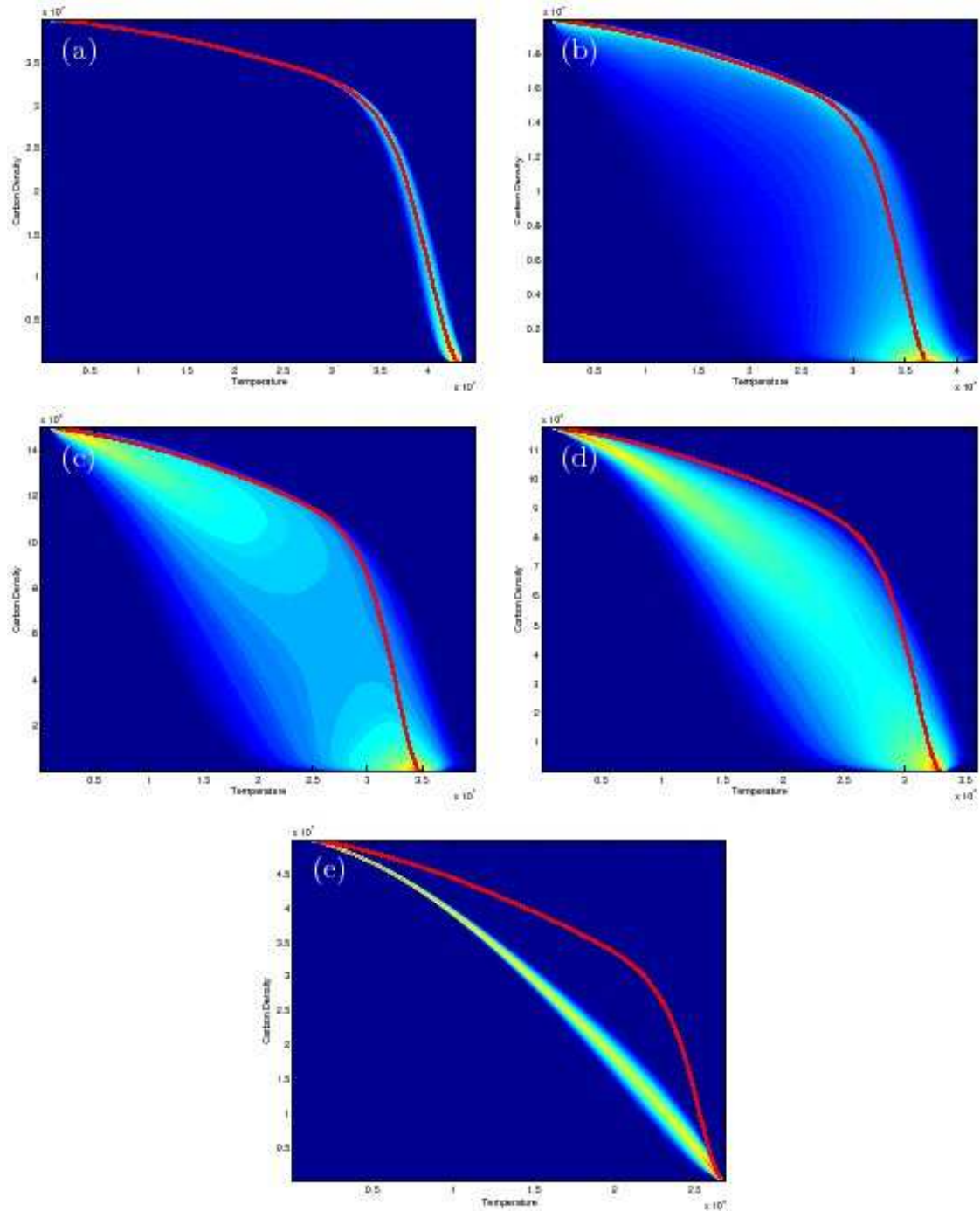}
\epsscale{1.0}
\caption{(a-e) Joint probability density functions of carbon density against
  temperature.  The red line denotes the
  laminar correlation.  Case (a) is very close to the laminar burning path.
  Cases (b), (c) and (d) show greater variation around the laminar path.  Case
  (e) present a collapse to a single curve disparate from the laminar case.  
  This is indicative of the competition between turbulent
  mixing and thermal diffusion; in case (a) thermal diffusion dominates the
  turbulence, in case (e) turbulent mixing dominates, and in the intermediate
  cases the effects of both processes are of comparable importance.}
\label{Fig:PdfRhoXcTemp}
\end{figure}

\begin{figure}
\centering
\epsscale{0.6}
\plotone{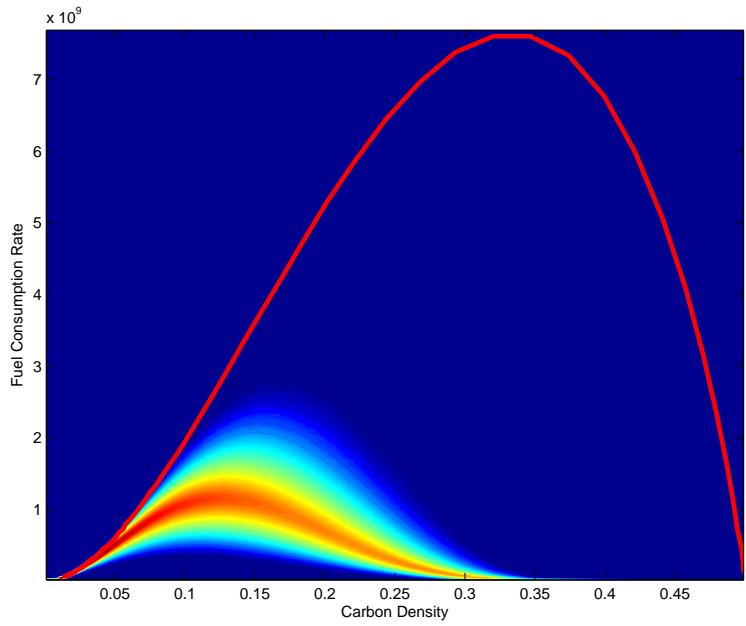}
\epsscale{1.0}
\caption{First moment with respect to fuel consumption rate of the joint
  probability density function of fuel consumption rate against carbon density
  for case (e); the other cases present similar behavior to
  figures~\ref{Fig:PdfRhoXcTemp} and~\ref{Fig:PdfTempFcrM1}.  This figure
  again shows the collapse to a burning path different from the laminar case
  indicating the dominance of turbulent mixing, and the lower local fuel
  consumption rates in this case, but also shows that the burning occurs at
  lower carbon mass fractions than in the laminar case.}
\label{Fig:PdfXcFcrM1}
\end{figure}

For densities near $2.35 \times 10^7$ g cm$^{-3}$ (Ka $\approx 3$),
turbulence begins to tear the flame, altering its width and speed, but
not completely disrupting the thin region where burning is going on.
This corresponds to the ``thin reaction zone'' regime of
\citet{Peters00}. By $1 \times 10^7$ g cm$^{-3}$ (Ka = 230), however,
the flame has been completely stirred and a qualitatively different
sort of distributed burning occurs. The width of the turbulent
flame is now much broader and it moves much faster. Turbulence
takes over from radiation as the dominant mode of energy transport,
even on small scales. Because of this, the ratio of heat diffusion to
composition diffusion approaches unity, i.e.~the Lewis number, which
previously was very large, approaches unity. This is clearly evident in
figure~\ref{Fig:PdfRhoXcTemp}, which shows the relation between
temperature and fuel concentration everywhere on the grid has
collapsed to a line corresponding to advective transport only.

\begin{figure}
\centering
\epsscale{0.6}
\plotone{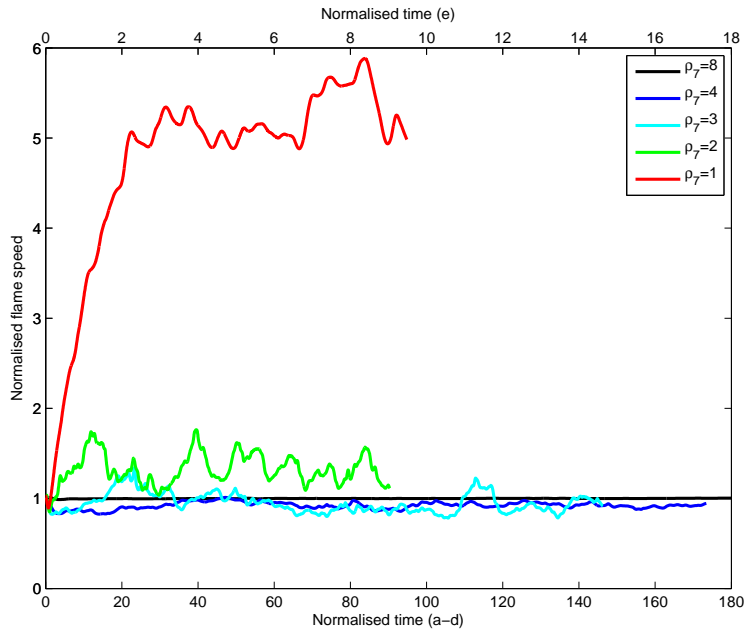}
\epsscale{1.0}
\caption{Normalized turbulent flame speeds.  The time scale has been
  normalized by the laminar burning time, i.e.~the time taken to burn one
  laminar flame width, specifically $l_L/s_L$.  Cases (a)-(d) use the lower time scale;
 case (e), which has a significantly different evolution, uses the upper time scale.
 The turbulent flame speed has been
  normalized by the laminar flame speed.}
\label{Fig:AllSpeeds}
\end{figure}

\begin{figure}
\centering
\begin{tabular}{cc}
\plottwo{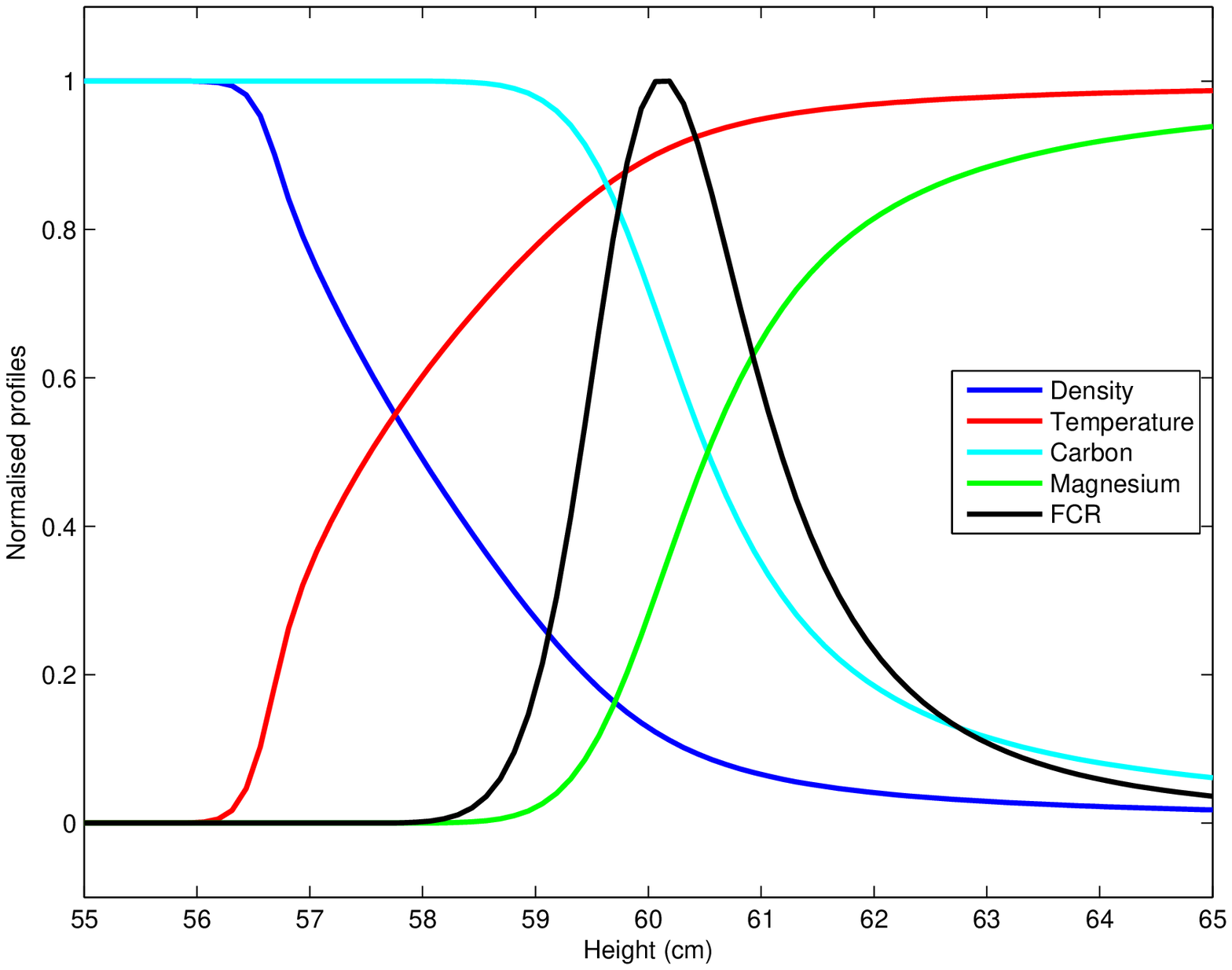}{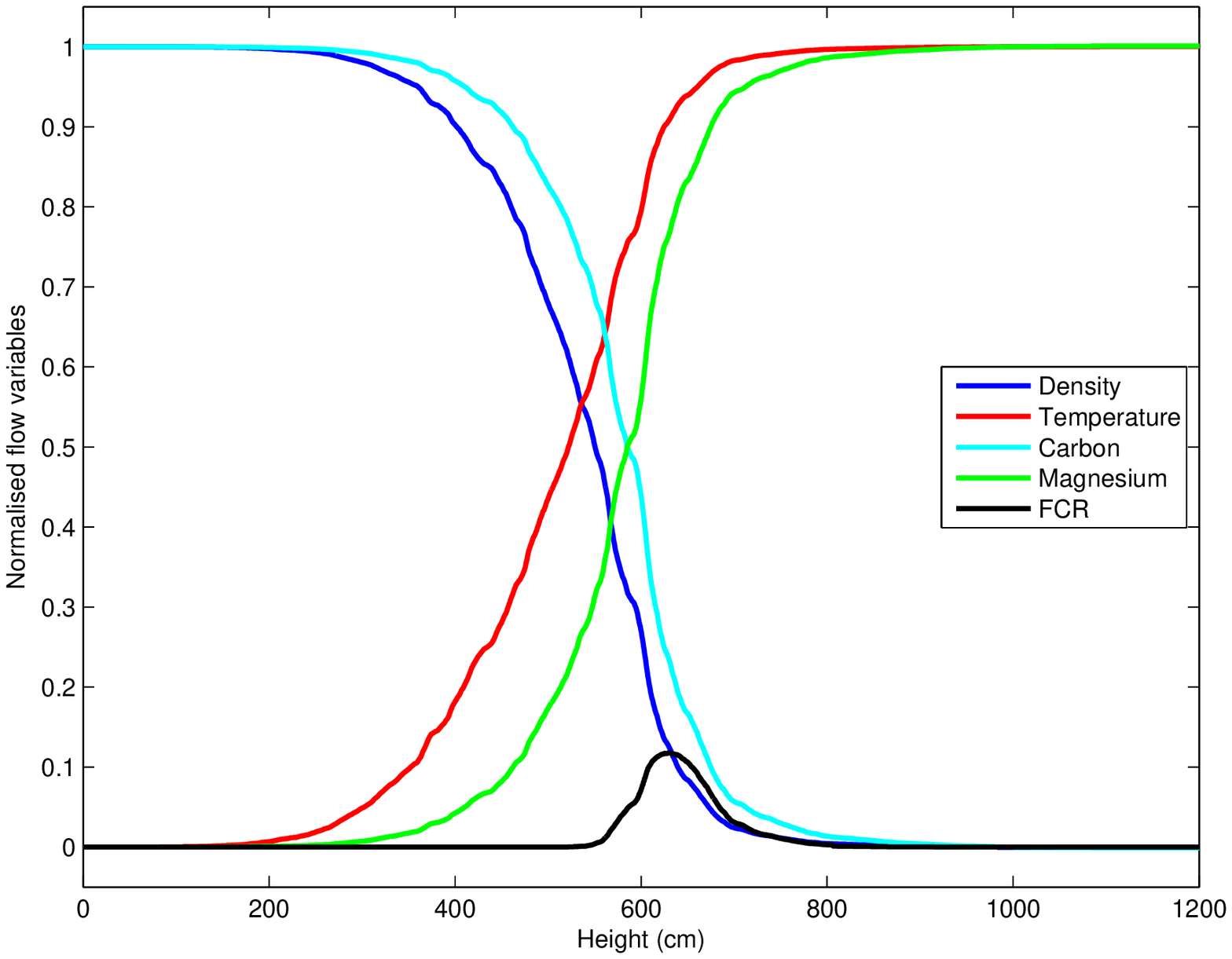}
\end{tabular}
\caption{Comparison of laminar (left) and turbulent (right) profiles in case (e).  In
  particular, note the disparate length scales between the two figures; the
  turbulent flame is around 60 times wider than the laminar flame.  Each flow variable has been
  normalized by the minimum and maximum values in the laminar flame, except
  for the velocity which has been normalized by the turbulent burning speed.
  Note the significant change in the shape of the profiles and the relative
  widths.  In the laminar case, thermal diffusion dominates, and so the
  temperature profile (and therefore density profile) is much wider than the
  species mass fraction profiles.  In the turbulent case, thermal diffusion is
  dominated by turbulent mixing and so the temperature, density and species
  mass fractions have similar profiles resembling hyperbolic tangents.
  Furthermore, the fuel consumption rate is greatly reduced in the turbulent
  case, but because of the greatly broadened flame width, the overall fuel
  consumption rate is enhanced.}
\label{Fig:FlameStructure}
\end{figure}

This is the first time supernova flames have been simulated in the
distributed regime in three dimensions, and our calculations may even
be a first in the combustion community as well. Terrestrial flames
with this degree of turbulence usually go out \citep{Peters00}. In a
supernova, with its long time scale and large size, extinction is
impossible until the star is completely disrupted. Still our results
at high Karlovitz number confirm burning in the distributed regime.
Assuming Damk\"ohler scaling \citep{Damkohler40}, the turbulent flame speed,
$s_T$, and its width, $l_T$, should obey the scaling relations
\begin{equation}
s_T \ = \ \sqrt{\frac{D_T}{\tau_{\rm nuc}^T}},
\end{equation}
\begin{equation}
l_T \ = \ \sqrt{D_T \tau_{\rm nuc}^T},
\end{equation}
for a turbulent diffusion coefficient, $D_T \sim u' \, l$, and nuclear time scale
$\tau_{\rm nuc}^T$.  It follows that
\begin{equation}
\frac{s_T}{s_L} \ = \ \left(\frac{u' \, l}{s_L \,
  l_L}\right)^{1/2} \ \left(\frac{\tau_{\rm nuc}^L}{\tau_{\rm
    nuc}^{T}}\right)^{1/2},
\label{Eq:TurbFlameSpeed}
\end{equation}
where $\tau_{\rm nuc}^L$ is the laminar nuclear time scale.
A key point is that the nuclear time scale is different in the laminar
and distributed cases. Because of the different distributions of
temperature and carbon abundance in the two cases
(figure~\ref{Fig:PdfRhoXcTemp}), the nuclear time scale is almost an
order of magnitude longer in the turbulent case. From the information
in Table 1, for case (e), $\tau_{\rm nuc}^L = l_L/s_L = 6.5
\times 10^{-4}$ s, while for the turbulent flame
(figure~\ref{Fig:FlameStructure}), $\tau_{\rm nuc}^T = l_T/s_T = 7.2
\times 10^{-3}$ s. This gives $s_T/s_L \approx 6.4$, approximately 15-20\% higher
than what
was calculated (figure~\ref{Fig:AllSpeeds}). The overestimate is
reasonable since the average turbulent diffusion coefficient is likely
to be somewhat less than that derived for the largest possible length
scale, i.e., $D_T < u' l$.

In the distributed burning regime with Karlovitz number greater than
ten and Damk\"ohler number less than one, i.e.~case~(e), the mixing is
dominated by turbulence and has a shorter time scale than the chemical
time scale.  Therefore, the overall reaction rate is limited by the
chemistry that can occur along the burning path prescribed by
figure~\ref{Fig:PdfRhoXcTemp}(e), and so the turbulent nuclear time
scale is a constant.  Together with the calculations presented
in~\cite{WoosleySciDAC08} and \cite{Woo08}, we believe that
case (e) has achieved the asymptotic value of $\tau_{\rm nuc}^T$. 

For larger integral scales than those simulated here, assuming that
$\tau_{\rm nuc}^T$ remains fixed, the turbulent speed, $s_T$, and
flame width, $l_T$, would both increase as $l^{2/3}$, as can be seen
from equation~(\ref{Eq:TurbFlameSpeed}) and the fact that $u' \propto l^{1/3}$. Such scaling
cannot continue indefinitely though, without eventually encountering
the limit $s_T \ltaprx u'$. The flame cannot move much faster than the
turbulence that carries it.

For our assumed turbulent energy and composition, the length scale
where $l^{2/3}$ scaling breaks down is predicted to be 
\begin{equation}
l_\lambda \ = \ \left(\frac{u'}{s_T}\right)^3 \, l,
\end{equation}
or, for $s_T = 1.95 \times 10^4$ cm s$^{-1}$ at $u' = 2.47 \times 10^5$
cm s$^{-1}$ and $l = 15$ cm, as calculated here, $l_\lambda \sim$ 300 m.
This compares reasonably well with the 155 m estimated by
\citet{Woosley2007} using a very simplified representation of the flame
structure.

The actual integral scale in the supernova is much larger even than
this large value. There, one expects a different sort of burning
reflecting some distribution of turbulently broadened flamelets with
characteristic scale $l_\lambda$ embedded in an overall flame brush as
large as the actual integral scale, $L^* \sim 10 $ km. Exploratory
calculations in progress suggest a highly variable flame width and
speed in that domain that may be conducive to spontaneous detonation.

As a practical consequence, the domination of turbulent transport at
low density, means one no longer needs to resolve the laminar flame scale.
It thus becomes possible to carry out meaningful simulations using 
much larger grid. In our next paper, we will demonstrate the validity 
of a simple sub-grid model for the turbulent transport and carry out 
three-dimensional calculations that take us into the unsteady burning regime. 

\acknowledgements

The authors thank F.~X.~Timmes for making his equation of state and
conductivity routines available online.
Support for A.~Aspden was provided by a Seaborg Fellowship
at Lawrence Berkeley National Laboratory under Contract No. DE-AC02-05CH11231.
Support for J.~Bell and M.~Day was provided by the SciDAC Program of the Office of
Advanced Scientific Computing Resarch of the U.S Department of Energy
under Contract No. DE-AC02-05CH11231.
At UCSC this research has been supported by the NASA Theory
Program NNG05GG08G and the DOE SciDAC Program (DE-FC02-06ER41438).
M.~Zingale was supported by a DOE/Office of Nuclear Physics Outstanding
Junior Investigator award, grant No.\ DE-FG02-06ER41448 to SUNY Stony
Brook.
The
computations presented here were performed on the Jaguar XT4 at ORNL as part of an INCITE award
and on the ATLAS Linux Cluster at LLNL as part of a Grand Challenge Project.

\newpage

\bibliographystyle{apj}
\bibliography{ms}

\clearpage

\begin{table}
\begin{tabular}{|l||c|c|c|c|c|}
\hline
Case & (a) & (b) & (c) & (d) & (e) \\
\hline
Fuel density ($\times10^7$g/cm$^3$) & $8$ & $4$ & $3$ & $2.35$ & $1$ \\
Ash density ($\times10^7$g/cm$^3$) & $5.74$ & $2.64$ & $1.91$ & $1.43$ & $0.52$ \\
\hline
Laminar flame width ($l_L$) & $5.00\times10^{-3}$ & $3.19\times10^{-2}$ & $7.24\times10^{-2}$ & $1.49\times10^{-1}$ & $2.31$ \\
Laminar flame speed ($s_L$) & $2.62\times10^{5}$ & $7.52\times10^{4}$ & $4.26\times10^{4}$ & $2.57\times10^{4}$ & $3.54\times10^{3}$ \\
Fuel Mach number ($s_L/c_F$) & $5.0\times10^{-4}$ & $1.6\times10^{-4}$ &  $9.7\times10^{-5}$ & $ 6.1\times10^{-5}$ & $ 1.0\times10^{-5}$ \\
\hline
Laminar peak FCR & $2.10\times10^{15}$ & $4.82\times10^{13}$ &$8.96\times10^{12}$ & $2.04\times10^{12}$ & $7.61\times10^{9}$\\
Laminar velocity change &  $1.03\times10^{5}$ & $3.85\times10^{4}$ & $2.45\times10^{4}$& $1.63\times10^{4}$ & $3.24\times10^{3}$\\
Laminar temperature change & $4.20\times10^{9}$ & $3.60\times10^{9}$ & $3.37\times10^{9}$ & $3.18\times10^{9}$ & $2.57\times10^{9}$\\
\hline
\end{tabular}
\caption{Laminar flame properties.}
\label{tbl:1}
\end{table}

\begin{table}
\begin{tabular}{|l||c|c|c|c|c|}
\hline
Case & (a) & (b) & (c) & (d) & (e) \\
\hline
Fuel density ($\times10^7$g/cm$^3$) & $8$ & $4$ & $3$ & $2.35$ & $1$ \\
\hline
Domain width ($L$) & $0.32$ & $2.0$ & $4.6$ & $9.5$ & $150$ \\
Domain height ($H$) & $1.28$ & $8.0$ & $18.4$ & $38.0$ & $1200$ \\
Integral length scale ($l$) & $0.032$ & $0.2$ & $0.46$ & $0.95$ & $15$ \\
Turbulent intensity ($u^\prime$) & $3.17\times10^{4}$ & $5.85\times10^{4}$ &
$7.72\times10^{4}$ & $9.83\times10^{4}$ & $2.47\times10^{5}$ \\
\hline
Gibson scale ($l_G$) & $1.8\times10^{1}$ & $4.3\times10^{-1}$ & $7.7\times10^{-2}$ & $1.7\times10^{-2}$ & $4.4\times10^{-5}$ \\
Karlovitz number $\sqrt{l_L/l_G}$ & $1.67\times10^{-2}$ & $0.274$ & $0.968$ & $2.96$ & $2.28\times10^{2}$ \\
\hline
\end{tabular}
\caption{Simulation properties.}
\label{Tab:SimProperties}
\end{table}

\clearpage

\clearpage

\clearpage

\clearpage

\clearpage

\clearpage

\clearpage

\clearpage

\clearpage

\clearpage

\clearpage

\clearpage

\clearpage

\clearpage

\end{document}